\documentclass{aa}
\usepackage{amsmath}
\usepackage{graphicx}
\usepackage[utf8]{inputenc}

\usepackage{txfonts}
\graphicspath{{Graficos/CV/}}

\begin{document}

   \title{Impacts on Ceres and Vesta: \\ Source regions, cratering, and fragmentation}
   \author{P. S. Zain\inst{1,2}\thanks{pzain@fcaglp.unlp.edu.ar}, 
          R. P. Di Sisto\inst{1,2} , 
           \and G. C. de El\'{\i}a\inst{1,2},
           }

   \offprints{Patricio Zain}
  \institute{Instituto de Astrof\'{\i}sica de La Plata, CCT La Plata-CONICET-UNLP \\                                                                                       
   Paseo del Bosque S/N (1900), La Plata, Argentina                                                             
   \and Facultad de Ciencias Astron\'omicas y Geof\'{\i}sicas, Universidad Nacional de La Plata \\                                                                           
   Paseo del Bosque S/N (1900), La Plata, Argentina 
                }

   \date{Received / Accepted}

%-------------------------------------------------------

\abstract
%context
{Ceres and Vesta are the largest members of the main belt (MB). They were visited by the NASA Dawn spacecraft between 2011 and 2018, which provided a great sample of detailed observations of the surface of both bodies.}
%aims
{We perform a study on the impacts on Ceres and Vesta. We aim to determine the size-frequency distribution (SFD) of impactors and to identify and quantify the contribution of each source region, as well as the craters produced and fragments ejected in these impact events. }
%methods
{We used a multipart collisional evolution model of the MB called \texttt{ACDC} (Asteroid Collisions and Dynamic Computation) that simulates the collisional evolution of the MB, which is split into six regions (namely Inner, Middle, Pristine, Outer, Cybele, and High-Inclination belts), according to the positions of the major resonances present there ($\nu_{6}$, 3:1J, 5:2J, 7:3J, and 2:1J).\ Furthermore, it includes the Yarkovsky effect as a dynamical remotion mechanism. We applied \texttt{ACDC} to Ceres and Vesta by keeping a record of all the bodies larger than $100$ m that hit them during $4$ Gyr. We performed 1600 simulations and,  for our analysis, selected the runs that provide the best fits with the SFD of the six regions of the MB and also those that are able to form the Rheasilva and Veneneia, the two large basins on Vesta. }
%results
{The six regions of the MB provide, to a greater or lesser extent, impactors on Ceres and Vesta. The Outer belt is the main source of impactors smaller than $10$ km on Ceres, providing more than half of the impacts, while the Middle belt is the secondary source. On Vesta, the relative impactor contribution of the Inner, Middle, and Outer belts is almost even. We are able to reproduce the craters larger than $100$ km in Vesta and identify two large depressions identified in Ceres as impact craters: one called Vendimia Planitia of $\sim 900$ km and a second one of $\sim 570$ km. As an outcome of these impacts, Ceres and Vesta eject fragments into the MB. We obtain fragmentation rates of tens of fragments larger than $1$ m per year for both bodies, to tens of fragments larger than $100$ m per million years for Vesta and a factor of $\sim 4$ greater for Ceres. We find that hundreds of bodies larger than $10$ km should have been ejected from Ceres and Vesta during their history. }
%conclusions
{}

\authorrunning{P. S. Zain et al.}
\titlerunning{Impacts on Ceres and Vesta: source regions, cratering and fragmentation}

\maketitle
\section{Introduction}

The Main Belt (MB) of asteroids is a vast region in our Solar System between Mars and Jupiter, from $\sim 2$ to $\sim 3.4 $ au from the Sun. In particular, Ceres and Vesta are the largest and most massive members in the MB. Both bodies were visited by the NASA Dawn space mission, which provided detailed observations of their surface and impact craters. These observations and this analysis changed some paradigms that constrain dynamical and physical models of Ceres and Vesta. Given the new observations by Dawn that reveal new and interesting data and the fact that Ceres and Vesta are the principal members of the MB, it is appropriate to develop a new collisional study on Ceres and Vesta. 

Vesta is one of the largest asteroids in the MB. It is located in the Inner belt, with a semimajor axis of $a =2.364$ au. Observations with Dawn have shown that Vesta is a triaxial ellipsoid with radii of $286.3 \times 278.6 \times 223.2$ km, a mean radius of $262.7$ km, and a mean density of $3456$ kg/$\text{m}^{3}$ \citep{Russell2012}. Vesta's most interesting features are two inmense impact basins located in the southern hemisphere that are as big as the asteroid itself: Rheasilvia and Veneneia, with diameters of $\sim 500$ km and $\sim 400$ km, respectively \citep{Thomas1997,Schenk2012}. The impacts that formed these basins are expected to be the source of howardite, eucrite and diogenite (HED) meteorites \citep{Desanctis2012a,McSween2013}. \cite{marchi2012} found that Rheasilvia was formed 1 billion years ago, in agreement with estimations on the age of the Vesta family \citep{Marzari1996}. Previous works calculated the size of the projectile that formed Rheasilvia. For example, \cite{deelia2011} estimated a projectile size of $\approx66$ km, while \cite{Jutzi2013} provide a similar estimate of $60-70$ km. The cratering production and chronology on Vesta has also been largely studied \citep{Schmedemann2014,marchi2012,Marchi2014,Vincent2014,OBrien2014,Roig2020}, and recently a global database of impact craters larger than $700$ m has been published \citep{Liu2018}. The impact events that formed these craters modified the surface of Vesta. In particular, hydrated material has been found in the Marcia crater, \citep{DeSanctis2015}, while the olivine present in the surface is assumed to be excavated from the mantle by impacts \citep{Ammannito2013,Turrini2016}. 

Ceres is the largest body in the main asteroid belt, with a semimajor axis of $a = 2.767$ au, it is located in the Middle Region of the MB. The detailed observations performed by the Dawn mission allowed to determine the real form of Ceres as a tri-axial ellipsoid of $483.1 \times 481.0 \times 445.9 $ km size with a mean radius of $469.7$ km  and a mean density of $2162$ $\text{kg}/\text{m}^3$ \citep{Russell2016}. Various studies about surface and crater morphology, surface composition and structure, water-ice deposits, and the implications of those observations in Ceres' surface suggest that Ceres is partially differentiated with an ice-rich upper mantle, a crust that is composed of a rock-ice mixture, and a rocky core \citep{Buczkowski2016, Bland2016, Hiesinger2016, Platz2016, Prettyman2017}. In fact, \cite{Prettyman2017} found that surface materials were processed by the action of subsurface water ice where it can survive for a billion years. By analyzing the morphology of large craters, \cite{Bland2016}  concluded that they are in general too deep and therefore inconsistent with viscous relaxation in a pure-ice layer, but they are made up of a mixture of rock, salts, and/or clathrates and   30\% to 40\% ice.  However, there are several shallow craters with limited viscous relaxation that may indicate spatial variations in subsurface ice content. On the surface, $\text{H}_2\text{O}$ ice was detected in isolated regions such as the Oxo crater \citep{Combe2016}, which is 10 km in diameter, and in bright deposits on the floors of ten craters \citep{Platz2016}. \cite{Desanctis2017} also found an organic-rich area where abundant ammonia-bearing hydrated minerals, water ice, carbonates, salts, and organic material were detected.  A very interesting feature was the observation of water vapor around Ceres \citep{Kuppers2014}, originating from localized sources that could be connected to comet or ice-rich asteroids impacts and/or cryovolcanism \citep{Ruesch2016}.

Cratering counting and analysis provided a global catalog up to a 1 km size,  regional chronologies, and studies of crater morphology \citep{Marchi2016, Hiesinger2016, gou2018, Otto2019, Nathues2017,Roig2020,Bottke2020}. However, observations have shown that the surface of Ceres lacks craters larger than $\sim 280$ km in diameter and it is depleted of craters down to $100-150$ km, which is incompatible with collisional models \citep{Marchi2016}. This shows that Ceres has gone through geological processes that obliterated those missing large craters over large timescales. However, \cite{Marchi2016} also provide evidence of a huge 800 km diameter depression, which may be a relict impact basin in a region called Vendimia Planitia, located on  Ceres' northern hemisphere, and this suggests two other depressions of $500$ and $800$ km. 

Recent works show that while Ceres and Vesta have not been targets of catastrophic collisions, they have been exposed to cratering impacts over the age of the Solar System \citep{deelia2011,marchi2012,gou2018}. A dynamical family has been found for Vesta \citep{Williams1979,Zappala1994}. However, detection techniques for identifying asteroid families did not find any Ceres family \citep{Milani2014}, and recent works explore different hypotheses and possibilities regarding this subject. In particular, \cite{Rivkin2014} suggest that the family-forming event would have excavated icy material from the mantle which would be significantly sublimated, while \citet{Carruba2016} argue that secular resonances with Ceres are able to deplete the population of objects in the proximity of this body. 

Recently, a new multipart collisional evolution model of the MB called \texttt{ACDC} (Asteroid Collisions and Dynamic Computation) was presented in  \cite{Zain2020}. It is a statistical code, based on the prescriptions of \citet{OBrien2005}, \citet{Bottke2005a}, \citet{deElia2007}, \citet{Morbidelli2009}, and \citet{Cibulkova2014} that simulates the collisional evolution of the MB, split into six regions bounded by the major resonances present in the MB, and it includes the action of the Yarkovsky effect and resonances as the mechanism that removes asteroids from the MB. 

This work is an attempt to look a little further into the collisional history of Ceres and Vesta. To do so, we used \texttt{ACDC} and studied the impactors that hit Ceres and Vesta, the source regions of the impactors, the cratering made in those events, and the fragments that were ejected from both bodies.

The paper is structured as follows. In Section 2 we briefly summarize the collisional evolution model. In Section 3, we present the cratering laws we used for Ceres and Vesta. In Section 4, we describe how we selected our runs for our analysis. We present our results for the impactors on Ceres and Vesta, their source regions, the cratering, and the fragmentation in Section 5. Finally, the conclusions of this work and the discussions regarding the results are presented in the last section.

 \section{Collisional evolution of the MB}

\subsection{Partition of the MB}

\begin{figure}[htp]
\centering
\includegraphics[width=8cm]{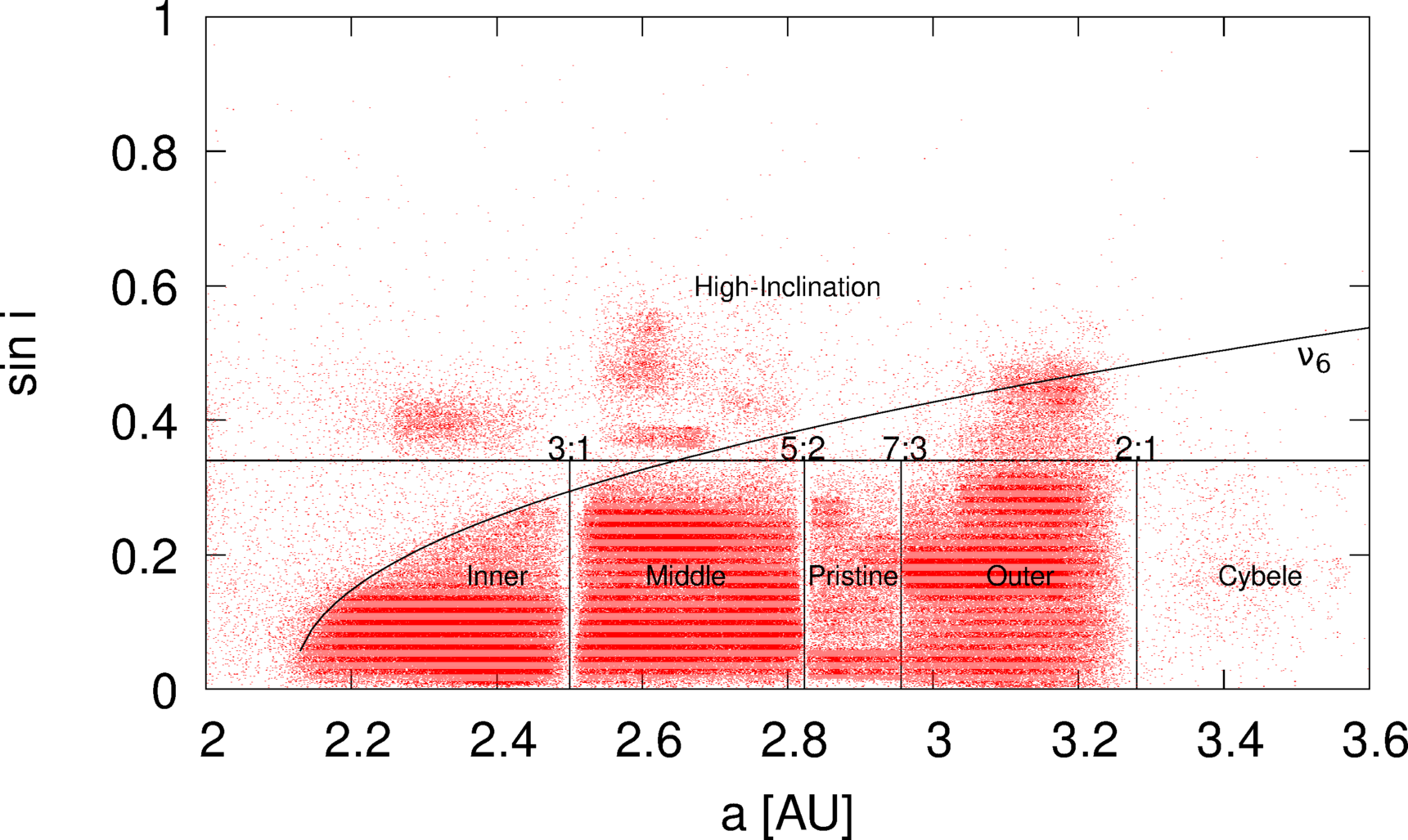}
\caption{Main asteroid belt, plotted in semimajor axis $a$ versus inclination $I$. The six defined regions are Inner, Middle, Pristine, Outer, Cybele, which are separated by the positions of the major resonances in the asteroid belt, and the High-Inclination region where $\sin i > 0.34$. The resonances considered are $\nu_{6}$, 3:1, 5:2, 7:3, and 2:1. The curve denoting the bond of the $\nu_{6}$ resonance was plotted according to \cite{Morbidelli2003}. Data were obtained from Minor Planet Center.}
    \label{fig:MB}
\end{figure}

In order to account for the different physical and dynamical properties in the different parts of the MB, we followed \cite{Cibulkova2014} and split the MB into six regions, or populations, separated by the major mean-motion resonances (MMR) with Jupiter and the $\nu_{6}$ secular resonance with Saturn. The MB and the locations of the MMR are plotted via the semimajor-axis versus inclination in Fig.~\ref{fig:MB}, using the online data from the Minor Planet  Center \footnote{\texttt{https://www.minorplanetcenter.net/iau/MPCORB/MPCORB.DAT.}}. 

The six regions are defined as follows: 

\begin{enumerate}
\item Inner belt: 2.1 AU < $a$ < 2.5 AU (3:1J);
\item Middle belt: 2.5 AU < $a$ < 2.823 AU (5:2J);
\item Pristine belt: 2.823 AU < $a$ < 2.956 AU (7:3J); 
\item Outer belt: 2.956 AU < $a$ < 3.28 AU (2:1J);
\item  Cybele belt: 3.28 AU < $a$ < 3.51 AU;
and\item  High-Inclination belt:  $\sin i $ > 0.34 ($i \gtrsim 20 \degr$) ($\nu_{6}$ secular resonance).
\end{enumerate}

The intrinsic probabilities of collisions  $P_{\textrm{imp}}$ and mutual impact velocities $v_{\textrm{imp}}$ between asteroids of the different regions were calculated by \cite{Cibulkova2014}, using the code written by \cite{BottkeGreenberg1993}. The mean values are listed in Table \ref{tab:ProbVimp}. We assume the MB to be in a collisional steady-state, so the probabilities and velocities will not change in time.  

\begin{table}
\caption{Intrinsic collisional probabilities $P_{\textrm{imp}}$ and mutual impact velocities $v_{\textrm{imp}}$ between bodies belonging to the different regions of the MB \citep{Cibulkova2014}.} 
\begin{center}
\begin{tabular}{|c c c|}
\hline
\hline
Populations & $P_{\textrm{imp}} \left(10^{-18} \text{km}^{-2} \text{yr}^{-1}\right)$ & $v_{\textrm{imp}} \left(\text{km }  \text{s}^{-1}\right)$ \\
\hline
\hline
Inner-Inner    & 11.98 & 4.34 \\
Inner-Middle   & 5.35  & 4.97 \\
Inner-Pristine & 2.70  & 3.81 \\
Inner-Outer    & 1.38  & 4.66 \\
Inner-Cybele   & 0.35  & 6.77 \\
Inner-High Inc.& 2.93  & 9.55 \\
\hline
Middle-Middle   & 4.91  & 5.18 \\
Middle-Pristine & 4.67  & 3.96 \\
Middle-Outer    & 2.88  & 4.73 \\
Middle-Cybele   & 1.04  & 5.33 \\
Middle-High Inc.& 2.68  & 8.84 \\
\hline
Pristine-Pristine & 8.97  & 2.22 \\
Pristine-Outer    & 4.80  & 3.59 \\
Pristine-Cybele   & 1.37  & 4.57 \\
Pristine-High Inc.& 2.45  & 7.93 \\
\hline
Outer-Outer    & 3.57  & 4.34 \\
Outer-Cybele   & 2.27  & 4.45 \\
Outer-High Inc.& 1.81  & 8.04 \\
\hline
Cybele-Cybele   & 2.58  & 4.39 \\
Cybele-High Inc.& 0.98  & 7.87 \\
\hline
High Inc.-High Inc.& 2.92  & 10.09 \\
\hline

\end{tabular}
\end{center}

\label{tab:ProbVimp}
\end{table}

\subsection{ Fragmentation laws}

Here, we present the model used to describe the outcome of a single collision between two bodies, which is based on the prescriptions of the \texttt{BOULDER} code \citep{Morbidelli2009}. The kinetic energy of the impact is the fundamental quantity that determines the outcome of the collision. In particular, we use the specific impact energy of the projectile $Q$, given by:
\begin{equation}Q=\frac{1}{2}\frac{m_{j} v_{\text{imp}}^{2}}{\left(m_{i}+m_{j}\right)},\end{equation}where $m_{i}$ and $m_{j}$ are the masses of the target and impactor, respectively. We assume that all bodies are spherical with uniform density $\rho$. We note that $v_{\text{imp}}$ is the mutual impact velocity and its values, depending on the source regions, are listed in Table \ref{tab:ProbVimp} \citep{Cibulkova2014}.

We compare the specific impact energy with the scaling law $Q_{\text{D}}^{*}$  and consider two impact regimes. Firslty, if $Q<Q_{\text{D}}^{*}$ , we have a cratering event. The target ejects fragments into space and a crater is created in the target's surface. Secondly, if $Q \geq Q_{\text{D}}^{*}$, we have a catastrophic disruption event, in which more than half of the total initial mass is dispersed in the form of fragments.

The largest surviving bodies after the impact are the largest remnant $M_{\text{LR}}$, and the largest fragment, $M_{\text{LF}}$. The expressions for both, as well as the slope $q$ of the fragments' SFD can be calculated as follows \citep{BenzAsphaug,Durda2007}:
\begin{equation}
M_{\text{LR}}=\begin{cases}
\left[-\frac{1}{2}\left(\frac{Q}{Q_{\text{D}}^{*}}-1\right)+\frac{1}{2}\right]\left(m_{i}+m_{j}\right), & Q<Q_{\text{D}}^{*} \text{ (cratering)}\\
\\
\left[-\frac{7}{20}\left(\frac{Q}{Q_{\text{D}}^{*}}-1\right)+\frac{1}{2}\right]\left(m_{i}+m_{j}\right), & Q \geq Q_{\text{D}}^{*} \text{ (catastrophic)}
\end{cases}
 \label{eq:MLR}
\end{equation}
\begin{equation}
M_{\text{LF}}=8\times10^{-3}\left[\frac{Q}{Q_{\text{D}}^{*}}\exp\left(-\left(\frac{Q}{4Q_{\text{D}}^{*}}\right)^{2}\right)\right]\left(m_{i}+m_{j}\right),
\label{eq:MLF}
\end{equation}
\begin{equation}
q=-10+7\left(\frac{Q}{Q_{\text{D}}^{*}}\right)^{0.4}\exp\left(-\frac{Q}{7Q_{\textrm{D}}^{*}}\right).
\label{eq:pendiente}
\end{equation}

We used the scaling law $Q^{*}_{\text{D}}$ derived by \cite{BenzAsphaug} for monolithic basaltic targets at  5 $\text{km s}^{-1}$ impact speeds. Although the compositions and dynamical properties in the different regions of the MB are diverse, finding appropriate scaling laws that account for these differences is still an open subject. Previous works explored different scaling laws. For example, \cite{Cibulkova2014} constructed new $Q^{*}_{\text{D}}$ functions for rubble piles, but found that the scaling law of \cite{BenzAsphaug} makes better fits with observed data. \cite{Bottke2005a} and \cite{Cibulkova2014} tested different scaling laws and concluded that laws much different from \cite{BenzAsphaug} cannot be used for the MB since they fail to reproduce the observed asteroid families.

\subsection{The \texttt{ACDC} code}
\label{sec:ColEVol}

In this work we used the \texttt{ACDC}  code. In this section, we only summarize the basic outlines of the \texttt{ACDC} code. For a full description of the construction and implementation of the model, the reader is refered to \cite{Zain2020}. 

The \texttt{ACDC} is a statistical and multipopulation code that simulates the collisional and dynamical evolution of the MB by evolving the incremental SFDs in time in each of the defined regions in 4 Gyr. The bodies are distributed in fixed logarithmic size bins. The collisional component is determined by the change in the number of bodies in each region of the MB due to the objects destroyed and fragments ejected in collisions, following and adapting the algorithms of previous collisional evolution works to the multipopulation case \citep{OBrien2005,deElia2007,Bottke2005a} . To do so, in each time-step, the \texttt{ACDC} calculates the number of collisions between all pairs of target-projectile bodies, located at the different regions of the MB, according to their intrinsic impact probabilities, while the occurrence of big impacts is determined by Poisson statistics. From this, the \texttt{ACDC} distributes the fragments created in each event in the different size bins, and it removes the bodies that are catastrophically disrupted. The values for the intrinsic collision probabilities and the impact velocities used in the calculations of the number of collisions and outcomes of the events are listed in Table \ref{tab:ProbVimp} \citep{Cibulkova2014}.

The dynamical component of the model is given by the combined action of the Yarkovsky effect and resonances as the mechanism that removes asteroids from the MB . This depletion affects the collisional evolution of the MB since fewer smaller bodies means fewer collisions with larger bodies. Following \cite{Cibulkova2014}, we modeled the dynamical depletion as an exponential decay with an associated timescale that is calculated for each region of the MB. The timescale depends on the size of the region and considers the mean values of the seasonal and diurnal variants of the Yarkovsky effect \citep{Peterson1976,Burns1979,Rubincam1995,Vokrouhlicky1998,Farinella1998b,Vokrouhlicky1999,ReviewYarko2015}. 

 The initial conditions in our simulations are the starting SFDs of the six regions of the MB, which were constructed as three-slope cumulative power laws by exploration of the free-parameter space with a large set of runs \citep{Zain2020}, and with the addition of the parental bodies of observed asteroid families in their respective size bins according to \cite{Cibulkova2014} and \cite{Broz2013}.

\section{Cratering on Ceres and Vesta}

Based on our model, it is possible to calculate all the impacts on Ceres and Vesta and the resultant crater distribution over the whole collisional evolution, that is 4 Gyr.  Without a doubt, Dawn's greatest achievement has been verifying that Ceres and Vesta, the two largest MB asteroids, are very different. The characteristics of an impact crater on an object are strongly dependent on its composition and structure. Therefore, the parameters of the scaling laws associated with such events should be treated with care and by considering the differences between the targets. In addition, Dawn's detailed observations obtained a large number of physical specifications and parameters associated with craters that must be included when applying the corresponding scaling laws. The new knowledge of Ceres and Vesta provided by Dawn allows one to improve the cratering scaling laws and therefore the determination of the cratering of these objects with respect to previous estimations.

\subsection{Cratering laws}

To calculate the apparent transient crater diameter $D_t$, produced by a projectile of diameter $d$, on both targets Ceres and Vesta, we used the scaling law from \cite{Holsapple2007}:
\begin{equation}
D_{\text{t}}=K_{1}\left[ \left( \frac{gd}{2v_{i}}\right)\left(\frac{\rho_{t}}{\rho_{i}}\right)^{2\nu/\mu}+K_{2}\left(\frac{Y}{\rho_{t}v_{i}^{2}} \right) \left(\frac{\rho_{t}}{\rho_{i}}\right)^{\nu \left(2+\mu\right)/\mu}  \right]^{-\mu/\left(2+\mu\right)}d
\label{eq:Crater}
\end{equation}
where $\rho_{t}$  is the target density, $g$ is its surface gravity, $Y$ is its strength, $\rho_{i}$ is the density of the impactor, and $v_{i}$ is the impactor velocity. The two exponents $\mu$ and $\nu$ and the constants $K_1$  and $K_2$ characterize the target material.  
The first term is a measure of the gravity of the target at the time of cratering and the second term indicates the importance of the strength of the target.  Thus, if the first term dominates over the second term, the crater forms under the gravity regime and generally corresponds to large events. On the contrary, if the second term dominates, it defines the strength regime which is relevant for small events  \citep{Holsapple1993}.  The transition impactor diameter ($D_{\text{l}}$) from one regime to the other is found by equaling the two terms of Eq. \ref{eq:Crater}. 

From our collisional evolution model, we have the mutual impact velocities $v_{imp}$ between asteroids of the different regions of the MB, as calculated by \cite{Cibulkova2014} (see Table \ref{tab:ProbVimp}). We calculated the real impactor velocity over a massive target by considering the  gravitational focusing by  
\begin{equation}
v_{i}=(v_{imp}^2 + v_{e}^2)^{1/2} 
\label{vi:Crater}
,\end{equation}
where $v_e$ is the target escape velocity. We are going to consider three possible impact angles:  a mean angle of  $45 \degr$ and two extremes $15\degr$ and $90\degr$. The effect of the impact angle on the scale cratering law is obtained by multiplying the impact velocity in Eq. \ref{eq:Crater} by $\sin \theta$.   

As mentioned, Ceres and Vesta have different compositions, densities, and surface properties. Therefore, the parameters of the scaling cratering law and the final crater size are treated separately.

\subsection{Ceres}

On the basis of the morphological characteristics of craters on Ceres,  \cite{Hiesinger2016} propose that they were formed in a target surface consistent with the presence of ice, and the outer shell is probably an ice-rock mixture. In regards to crater morphology, \cite{Hiesinger2016}  found that craters on Ceres are indistinguishable from craters on mid-sized icy satellites of Saturn. 
Thus, following the analysis by \cite{DiSisto2013} based on the scaling laws for ice obtained by \cite{Kraus2011}, for Ceres, we consider the parameters corresponding to an ice surface such as the Saturnian satellites. Then, in Eq. \ref{eq:Crater}, $\mu=0.38$, $\nu=0.397$, and $K_1 = 1.67$. We assume the constant $K_2 = 0.8$  for cold ice from Holsapple's cratering theory\footnote{Web page   http://keith.aa.washington.edu/craterdata/scaling/theory.pdf Accessed June, 2020}. The strength $Y$ of the target is unknown for Ceres since it is a quantity that has to be directly measured. Instead, one could chose a value by considering an analogous material such as ice or a mix of rock and ice. However, \cite{Hiesinger2016} estimated the strength-to-gravity transition crater diameter $D_{\text{l}} = 1.75$ km  and this is directly related to the strength of the target. The value of $Y$ for which we obtained $D_l$, similar to the value from  \cite{Hiesinger2016}, is  $Y = 4 \times 10^6$ $\text{dyn}/\text{cm}^2$ which could correspond to a mix of rock and ice.
 
Eq. \ref{eq:Crater} corresponds to the transient diameter of a crater which scales to a simple crater. However, above a certain threshold size,  a simple crater  collapses because of gravitational forces, finally leading to complex craters. The simple to complex transition diameter $D_{\text{sc}}$ depends on the target properties and is an observable quantity. Based on observations of floor-fill material and on the inflection in the depth-diameter curve, \cite{Hiesinger2016}  found that $D_{\text{sc}}$ is at about $7.5$ to $12$ km. Also  $D_{\text{sc}}$  can be calculated with the treatment in \cite{Kraus2011}, where the simple-to-complex transition diameter is scaled at the value observed on Ganymede by the following relation:
\begin{equation}
D_{\text{sc}} = \frac{2g_{\text{g}}}{g}, 
\label{eq:dsc}
\end{equation}
where  $g_{\text{g}}$ is the surface gravity of  Ganymede (equal to $1.428 \text{m}/\text{s}^2$), and $g = 0.28 \text{m/s}^2$ is the surface gravity of Ceres. 
With this equation, $D_{\text{sc}} = 10.2$ km, which is in the range of the agreement with \cite{Hiesinger2016} observations. 
Following \cite{Kraus2011}, the final crater diameter $D_{\text{f}}$ for $D_{\text{t}} > D_{\text{sc}}$ is given by
\begin{equation}
\frac{D_{\text{f}}}{D_{\text{t}}} = (1.3 k)^{1/(1-\eta)} \left(\frac{D_{\text{t}}}{D_{\text{sc}}}\right)^{\eta/(1-\eta)}, 
\label{cc}
\end{equation}
where $k=1.19$ and $\eta=0.04$.
Then the final crater size can be obtained by 
\begin{xalignat}{4}
D = (1.3\, k)D_{\text{t}}  &&\text{for} && D_{\text{t}} \leq D_{\text{sc}}/1.3\,k,  \nonumber \\
D = D_{\text{f}}   &&\text{for} &&  D_{\text{t}} > D_{\text{sc}}/1.3\,k.
\label{dcrater}
\end{xalignat} 
For continuity in the transition from simple to complex craters, we used the factor $1.3\, k$, from transient to final simple craters according to  \cite{Marchi2011}. 

\subsection{Vesta}

Dawn observations confirmed that Vesta is differentiated, with an iron core, a mantle, and a basaltic crust \citep{Russell2012}. Its composition and the basaltic surface leads to the use of the scaling cratering law, corresponding to wet soil and the rock of \cite{Holsapple2007}, which was also previously used by other authors \citep{deelia2011,Marchi2014}. Therefore, in Eq. \ref{eq:Crater},  $\mu=0.55$, $\nu=0.4$, and $K_1 = 0.93$ from \cite{Holsapple2007} and the constant $K_2 = 0.8$ for hard rock from Holsapple 2020\footnote{Web page   http://keith.aa.washington.edu/craterdata/scaling/theory.pdf Accessed June, 2020}.  For the value of the strength $Y$, we used the typical one for  rock:  $Y = 1.5 \times 10^8$ $\text{dyn/cm}^2$. With this value of $Y$, we obtained a the strength-to-gravity transition crater diameter $D_{\text{l}} \sim 25$ km, which is in agreement with the value obtained by \cite{Vincent2014} of $D_{\text{l}} = 22.3$ km. 

The simple to complex transition diameter $D_{\text{sc}}$ for Vesta was determined by \cite{Vincent2014} by analyzing the depth-to-diameter variations over the whole surface. They found $D_{\text{sc}} = 38$ km and we use this value in our calculations.   

The final crater diameter was obtained in a similar way as for Ceres, but following  \cite{Marchi2011} whose expressions are the ones that are suitable for the composition of Vesta: 
\begin{xalignat}{4}
D = 1.3 D_{\text{t}}  &&\text{for} && D_{\text{t}} \leq D_{\text{sc}}/1.3,  \nonumber \\
D = 1.4 D_{\text{t}}^{1.18}/ D_{\text{sc}}^{0.18}   &&\text{for} &&  D_{\text{t}} > D_{\text{sc}}/1.3.
\label{dcratervesta}
\end{xalignat}

\section{Selection of runs}

\subsection{First selection of runs: Fit with MB observed data}

\begin{figure}[h]
    \centering
\includegraphics[width=8cm]{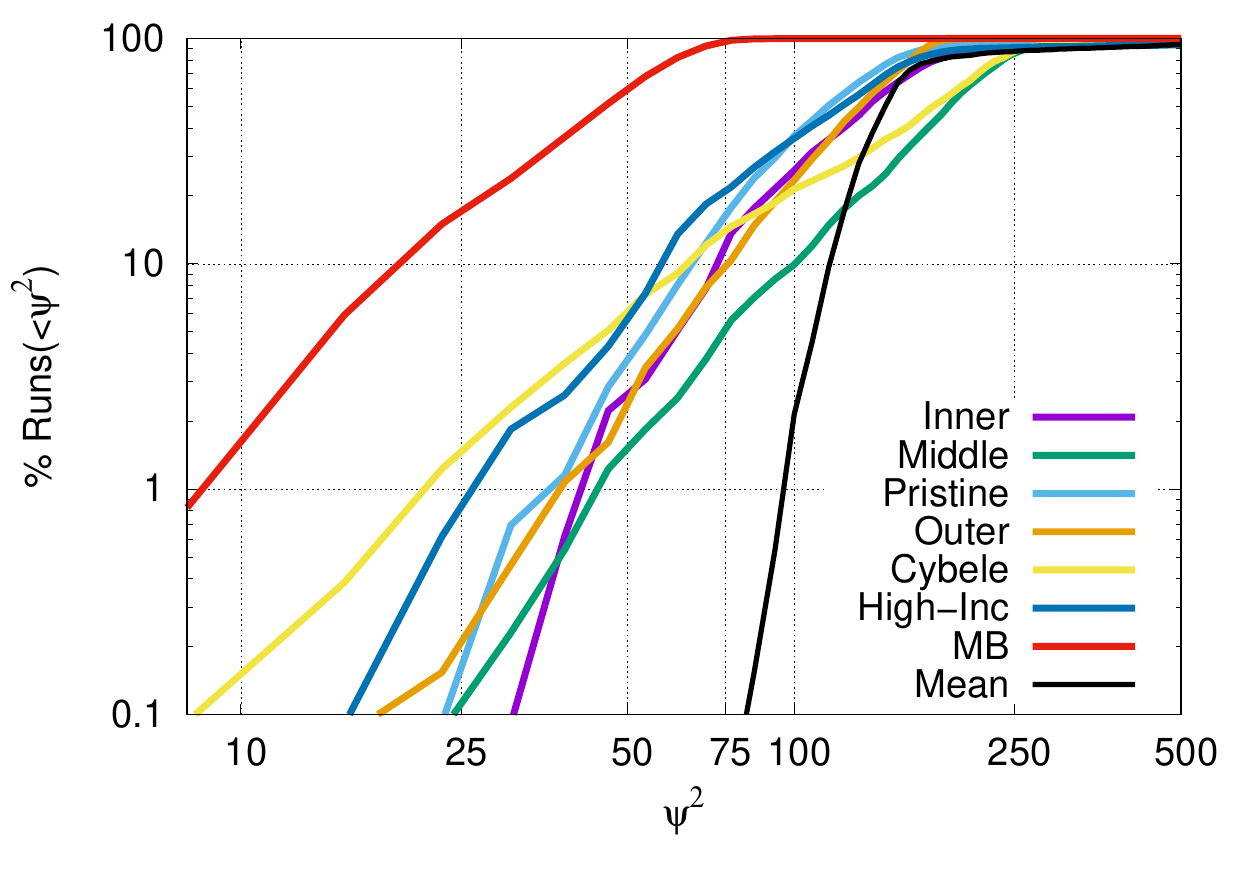}
    \caption{Cumulative percentage distribution of $\psi^{2}$ metrics. We plotted $\psi^{2}$ for the six regions of the MB, including a metric for the global MB $\psi^{2}_{\text{MB}}$ and the average  $\psi^{2}_{\text{MEAN}}$  metric.}
    \label{fig:Metric}
\end{figure}
\begin{figure*}[ht]
    \centering
\includegraphics[width=18cm]{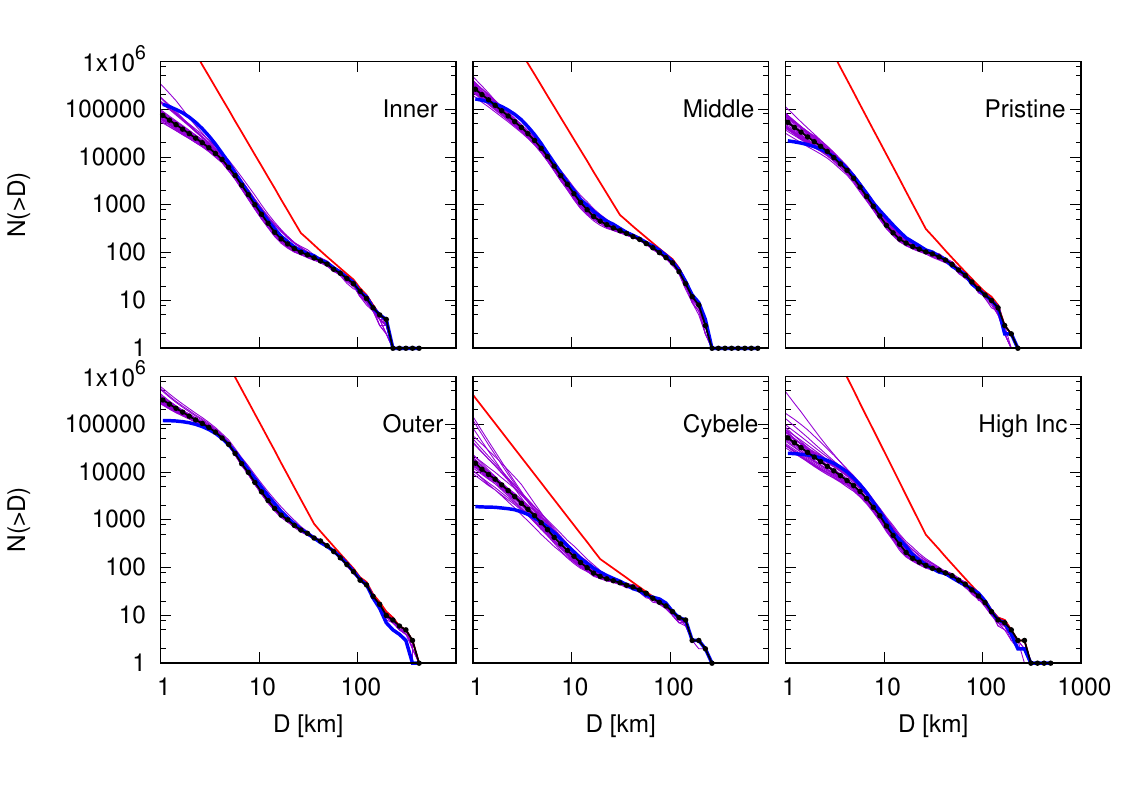}
    \caption{SFDs of the six regions of the MB. The first selection of runs and their median are plotted in purple and black lines, respectively. The observed SFDs are plotted in blue and the initial SFDs are plotted in red. }
    \label{fig:FinalSFD}
\end{figure*}

One of the main characteristics of our collisional evolution model is that it is highly stochastic, as it treats big impacts as random events using Poisson statistics. So, runs using different random seeds may produce much different results. We searched for runs that produce good fits in the individual regions and in the global MB. To do so, we developed a large set of runs and then we used observational constraints to select the best ones and interpret the results statistically. 

To obtain a quantitative measure of how good a simulation reproduces observational data, we followed the procedure described by \cite{Zain2020}, inspired by \cite{Bottke2005b} and \cite{Cibulkova2014}. The metric used to determine the goodness of fit between the observed size distribution of a region of the MB ($N_{\text{obs}}$) and the simulation results ($N_{\text{sim}}$) is a hybrid $\psi^{2}$ test:
\begin{equation}
\psi^{2}=\sum_{i}\left(\frac{N_{\text{sim}}\left(>D_{i}\right)-N_{\text{obs}}\left(>D_{i}\right)}{\sigma_{i}}\right)^{2},
\end{equation}
where the summatory extends over a range of diameters  $\sim$ 1 km - 250 km \citep{Cibulkova2014}. The uncertainties are given by $\sigma_{i}=0.1 N_{\text{obs}}\left(>D_{i}\right)$. For each run, we calculated the individual metrics $\psi^{2}_{r}$  for the six regions and also a mean metric $\psi^{2}_{\text{MEAN}}$ by averaging the metrics of the six regions.

We performed 1600 runs with \texttt{ACDC}. The first selection of runs was made by sorting the ones that make better fits with observed data, as it was described in detail in \cite{Zain2020}. In particular, we selected the runs that give $\psi^{2}_{\text{MEAN}}< 100$. 

The cumulative distributions of the  $\psi^{2}$ metrics of the six regions, along with the global MB  $\psi_{\text{MB}}^{2}$  and the averaged metric $\psi_{\text{MEAN}}^{2} $ of the performed runs, are plotted in Fig. \ref{fig:Metric}. We find that all of our runs produce good fits with the global MB. In fact, all of our runs produce $\psi_{\text{MB}}^{2} $ smaller than 75. The individual metrics give a wider range of values. The lowest metrics of the individual regions lie between 8 and 30. For our analysis, we selected the 26 runs that give mean metrics $\psi_{\text{MEAN}}^{2}$ smaller than 100, which represent $\sim2\%$ of the total. 

The resulting SFDs of the first selection of runs of the six regions of the MB, along with the median SFDs are plotted in Fig. \ref{fig:FinalSFD}. The observed size-frequency distributions of the different regions of the MB were constructed by \cite{Cibulkova2014} using observational data from the WISE satellite \citep{Masiero2011} and the AstOrb catalog \citep{Tedesco2002}. In general terms, we see that our selected runs provide very good fits with the observed data for the six regions and the global MB. As is discussed in \cite{Zain2020}, the main discrepancies are located in the small end of the Inner belt and the large end of the Outer belt, and they will be addressed in future research.

\subsection{Second selection of runs}
\begin{figure}[h]
\centering
\includegraphics[width=8cm]{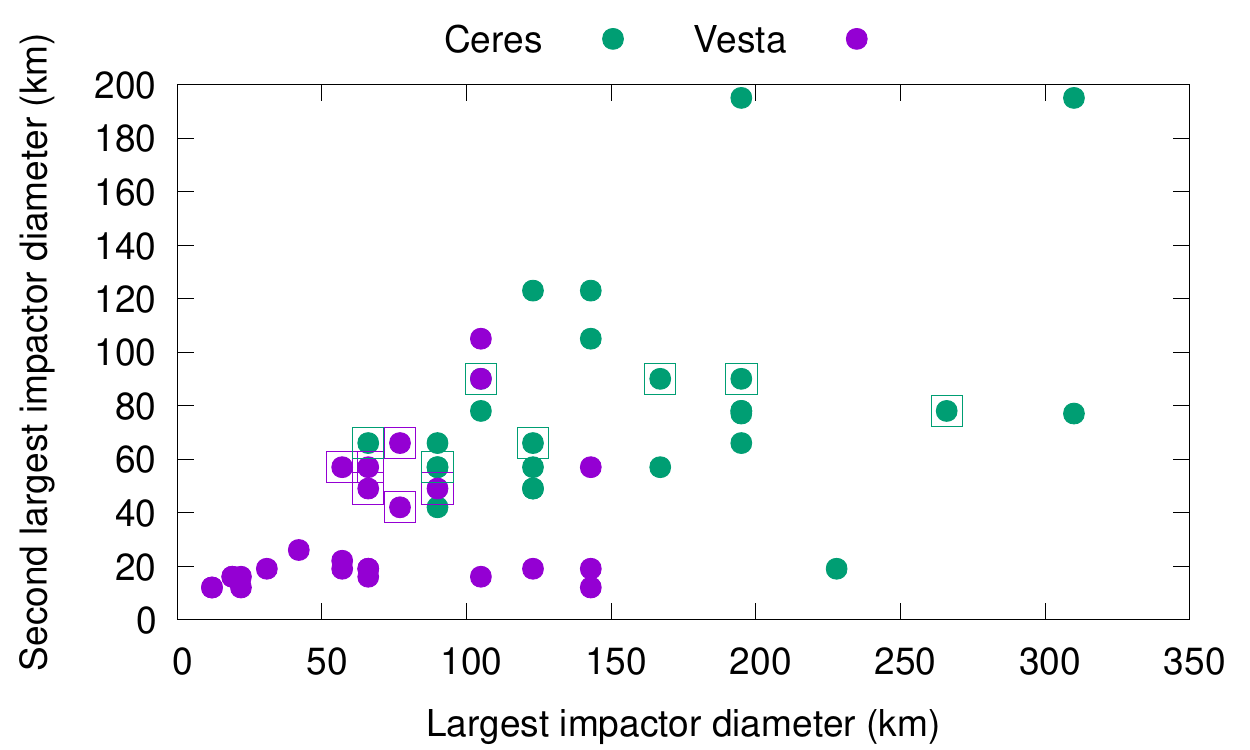}
\caption{Largest and second largest impactor on Ceres (green) and Vesta (violet) in the first set of selected runs. The squared dots indicate the impactors that are capable of forming the Rheasilvia and Veneneia craters on Vesta, and the corresponding pair of the largest impactors for Ceres. }
\label{fig:PySimp}
\end{figure}

We selected 26 runs by sorting the runs that make a better fit with observed data in the MB. However, since we are focusing on the collisional history of Ceres and Vesta, a second restriction must be made regarding the largest impacts that hit both bodies. We plotted in Fig. \ref{fig:PySimp} the pair or largest and second largest impactor that hit Ceres and Vesta in the first set of runs. Since our model treats big impacts as random stochastic events, the first selection of runs provide a wide variety of the largest impactors in both bodies. For Vesta, we obtain the largest impactors in the range from $12$ km - $143$ km, while for Ceres we obtain the largest impactors in the range from $66$ km - $310$ km. 

\begin{figure}[h]
\centering
\includegraphics[width=8cm]{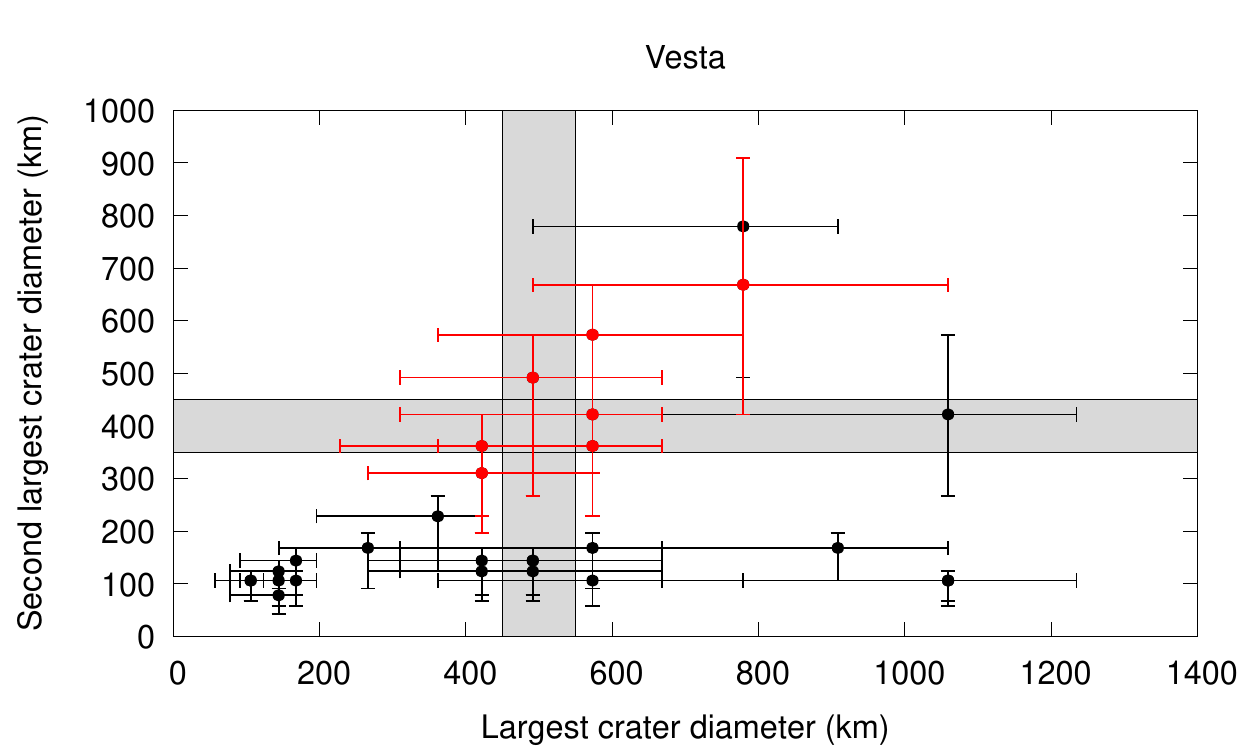}
\caption{Largest and second largest crater on Vesta in the first set of selected runs. The points denote an impact angle of 45\degr, while the error bars denote the angles of 15\degr \ and 90\degr. The shaded areas indicate a diameter of $500 \pm 50$ km and $400 \pm 50$ km for the largest and second largest crater diameter, respectively. We looked for runs that, with a right combination of impact angles, are able to form both large basins. The runs we selected for our analysis are plotted in red. }
\label{fig:PyScrat}
\end{figure}

However, not all of these runs are of interest to us since we are interested in studying Ceres and Vesta in particular. Thus, we restrained our analysis only to the runs that are capable of forming the Rheasilvia and Veneneia basins on Vesta, which have diameters of $\sim500$ km and $\sim400$ km, respectively \citep{Schenk2012}. To do so, we plotted in Fig. \ref{fig:PyScrat} the largest and second largest craters on Vesta, produced by the largest impactors shown in Fig. \ref{fig:PySimp}, with the scaling laws stated earlier in this paper. As the final crater diameter depends on the impact angle, we considered the most probable impact angle of  45\degr and also the following two extreme impact angles: 15\degr\  and 90\degr. Thus, for our further analysis, we selected the runs that, with a right combination of impact angles, are able to produce both the Rheasilvia and Veneneia basins on Vesta. The seven runs that fulfill these restrictions have largest impactor sizes between $57-77$ km for Rheasilvia and $42-66$ km for Veneneia, and one special case of $105$ km and $90$ km sizes with low impact angles. They are plotted as squared dots in Fig. \ref{fig:PySimp},  while the respective pair of the largest impactors in Ceres in this second selection of runs are plotted as squared green dots. 

\section{Results}
\subsection{Impactors - Source regions}

Here we focus on what we can learn about the asteroids that hit Ceres and Vesta during the history of the Solar System. To do so, we look forward to determine what asteroids hit Ceres and Vesta and where did these asteroids came from.  In particular, we aim to determine the source region of the impactors and a median size-frequency distribution.

\begin{figure*}[h]
\centering
\includegraphics[width=8cm]{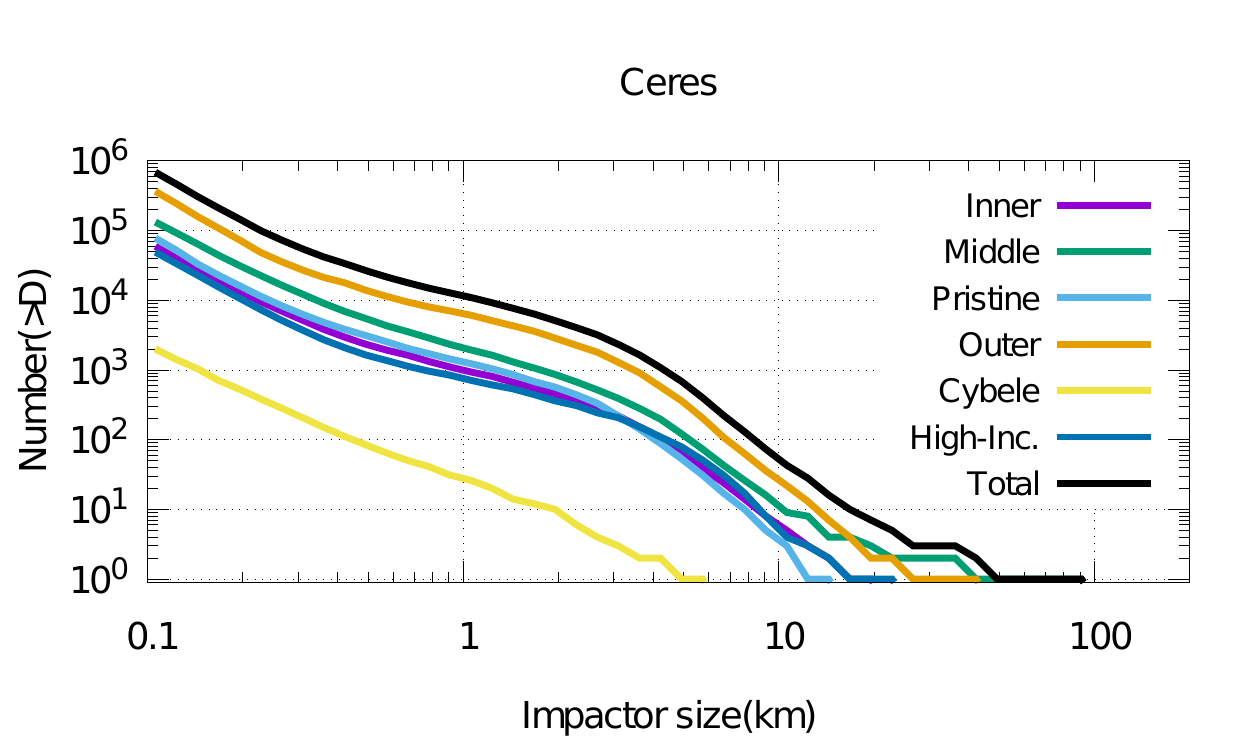}
\includegraphics[width=8cm]{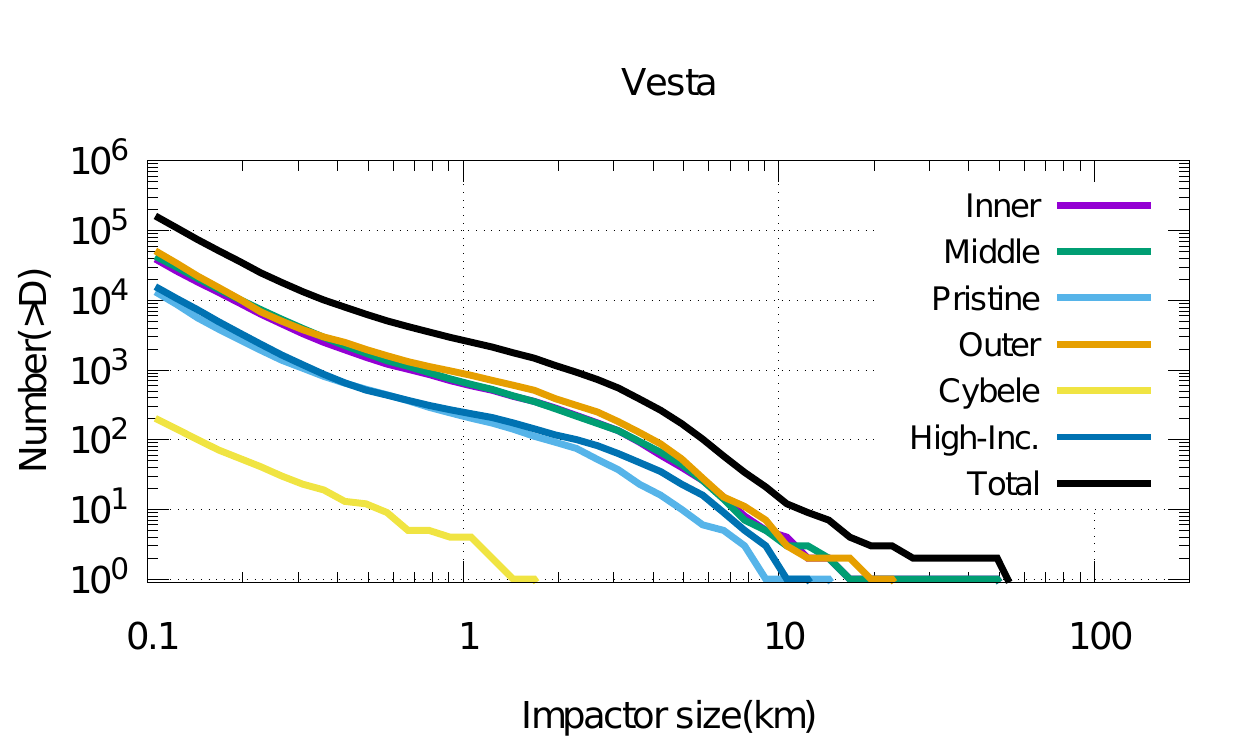}
\includegraphics[width=8cm]{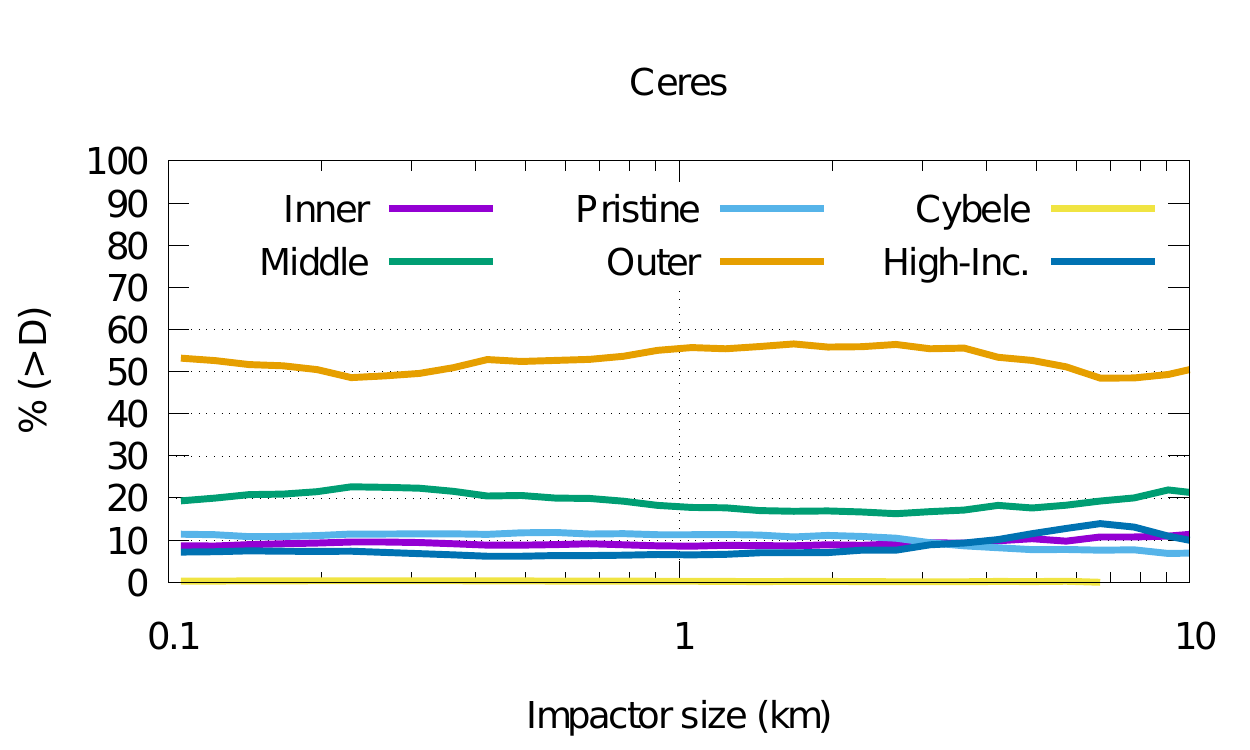}
\includegraphics[width=8cm]{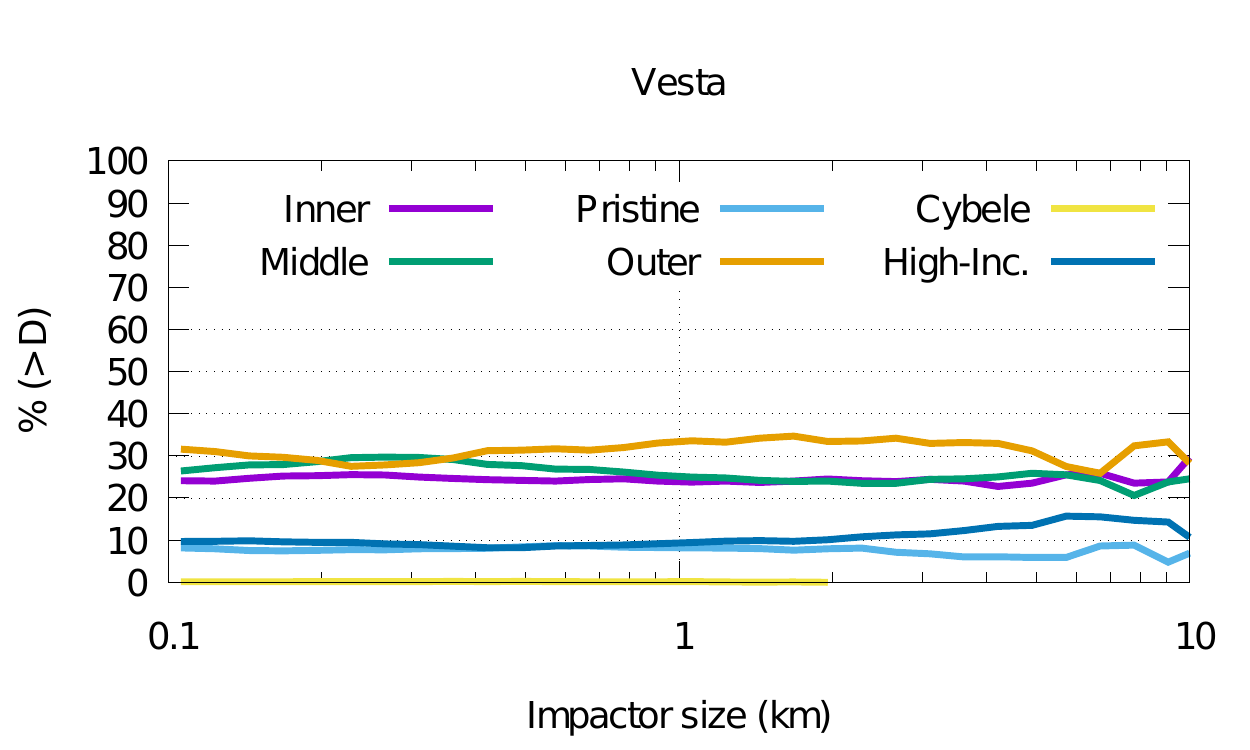}
\caption{Top: Median SFD of the impactors on Ceres and Vesta, coming from the different source regions of the MB, derived from the second selection of runs. The large end of this SFD, when $D>10-20$ km, must be handled with care due to the high stochasticity of big impact events and small numbers. Bottom: Relative contribution from each region of the MB to the impactors in the range $0.1-10$ km, with respect to the total cumulative number of impacts.}
\label{fig:Impactors}
\end{figure*}

The median SFDs of impactors coming from the different source regions on Ceres and Vesta, calculated with the second selection of runs, are plotted in Fig. \ref{fig:Impactors}. The large end of these SFDs, specifically when $D > 10-20$ km, must be handled with care as the results in the individual runs may show differences with respect to the median. This happens for two main reasons. The first reason is the small numbers, given that the number of impactors in the large end coming from the different regions is smaller than $\sim 10$ and decreases to unity in the case of the largest impactors. Second, given that the large impacts are highly stochastic as the occurrence is given by Poisson statistics, these individual projectiles in the different runs could come from any of the six regions of the MB. We limit the discussion regarding the source regions of impactors in the range of $0.1-10$ km, so we ensured that there were at least ten impacts and that the median distributions were statistically reliable. We plotted, in the bottom part of Fig. \ref{fig:Impactors}, the relative contribution of each region to the impactors, represented as the percentages with respect to the total cumulative number of impacts. At this point, we would like to note the following two things: the six regions of the MB contribute to, a greater or lesser extent, the impactors on Ceres and Vesta, and the relative contribution remains similar throughout the size range considered.

In the case of Ceres, we see that the Outer belt is clearly the main source of impactors smaller than $10$ km. In fact, the Outer belt provides $\sim56\%$ of impactors in the size range considered. The second main source is the Middle belt, the region where Ceres is located, which provides $\sim 20\%$ of impactors in the kilometer range. The contribution of the remaining regions is much lower. In fact, the Inner, Pristine, and High-Inclination regions provide a nearly even proportion of impactors, between $\sim 5\%-10\%$.  We see that the curves of the SFD from the Inner, Pristine, and High-Inclination impactors overlap in most size ranges. Finally, we see that the total contribution from the Cybele belt is negligible, being approximately two orders of magnitude lower than the rest. 

The fact that the Outer belt is the main source of impactors on Ceres could have implications on its observed surface properties. The dominant taxonomic class of asteroids in the Outer belt is C-type \citep{Demeo2013}. This class is known to have primitive material as they are dark, carbonaceous objects and volatile‐rich with a flat spectrum in the visible and infrared, with Ceres matching this C-type \citep{Demeo2009}. They  are associated with carbonaceous chondrite meteorite groups MI and MC \citep{Marchi2019}. Dawn observations and investigations revealed subsurface ice \citep[e.g.,][]{Bland2016, Prettyman2017} and isolated surface ice exposures \citep{Combe2016, Platz2016}.  In particular, \cite{Platz2016} detected bright deposits on the floors of ten craters, one of them corresponding to water ice. \cite{Desanctis2017} also found an organic-rich area where abundant ammonia-bearing hydrated minerals, water ice, carbonates, salts, and organic material were detected.  The presence of water ice and organic material on the surface of Ceres seems to be connected with aqueous alteration processes and interior evolution \citep{Prettyman2017, Desanctis2017}; however, the fact that the main impactors on Ceres are C-type asteroids could suggest that at least a certain proportion of volatile and carbonaceous materials could come from collisions received from the outer region. Considering that Ceres was formed or passed the great majority of its life in the MB, those C-type asteroids collisions could even had provided a portion of the now subsurface water ice.  Moreover, the observation of water vapor around Ceres \citep{Kuppers2014} could be due to the sublimation of ice from recent impacts of asteroids with a water ice content.

In the case of Vesta, which is located in the Inner belt, we find that the contribution is almost even between the Inner, Middle, and Outer belts. In fact, the mentioned regions provide $\sim24\%$, $\sim26\%$, and $\sim 32\%$ of impactors, respectively, in bodies smaller than $10$ km. We can see that the impactor SFDs from the mentioned regions overlap in most size ranges, with a slight majority of impactors from the Outer belt in the range from $1$ km-$4$ km. In second place, the Pristine and High-Inclination SFDs overlap in impactor sizes smaller than 1 km and provide $\sim$10\% each, but the High-Inclination contribution is slightly larger than Pristine in the range from $2$ km - $10$ km.

\subsection{Cratering}

In the previous section we discussed the impactors that hit Ceres and Vesta in the simulations we performed. Here we discuss the craters that result from these impact events. 

\begin{figure}[htp]
        \centering
                \includegraphics[width=8cm]{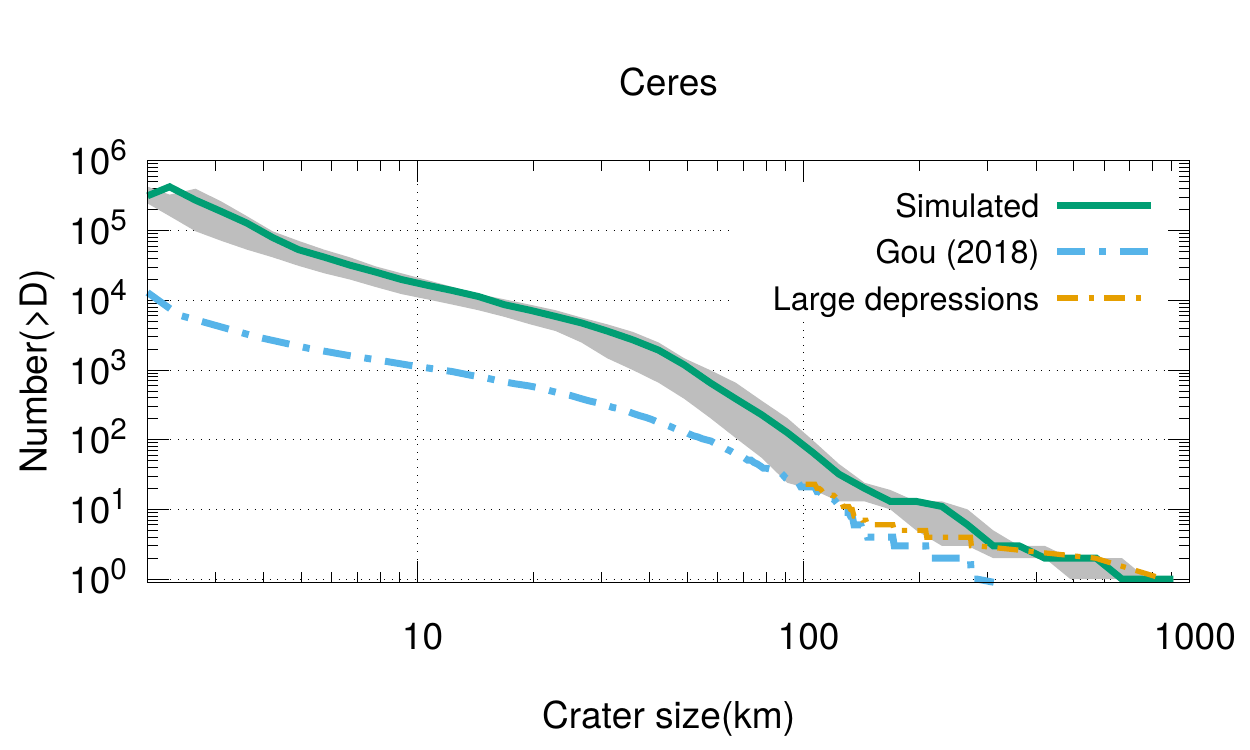}
                \includegraphics[width=8cm]{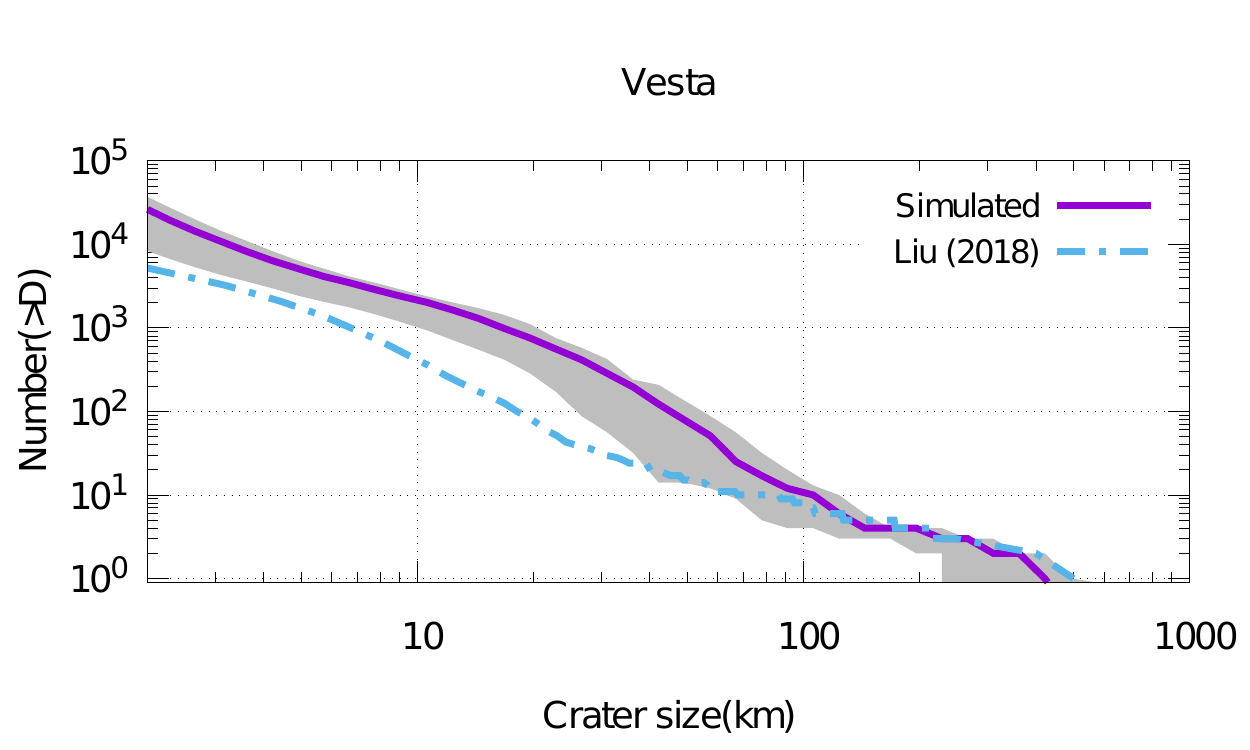}
        \caption{Simulated and observed crater SFD for Ceres (Top) and Vesta (Bottom) in the run that produces the best fit. The solid line represents an angle of $45\degr$ for all impacts, while the top part of the shaded area indicates an impact angle of $90\degr$ and the bottom part and angle of $15 \degr$. The crater counts were taken from \cite{Liu2018} and \cite{gou2018}. In the case of Ceres (Top), a second line is observed, which is made of the crater catalog of \cite{gou2018} with the addition of two large depressions suggested by \cite{Marchi2016}. }
        \label{fig:CraterTotalCV}
\end{figure}

We compared the craters obtained in our simulations with the cataloged craters published by \cite{Liu2018} for Vesta and \cite{gou2018} for Ceres. The global catalog of craters on Ceres by \cite{gou2018} accounts for  $29\,219$ craters with a diameter $D \geq 1$ km and it covers approximately $98 \%$ of Ceres' surface. For Vesta, \cite{Liu2018} built a global database containing $11\,605$ craters with $D \geq 0.7$ km by visual crater identification and a mathematical shape determination with Dawn images at a resolution of $\sim 60$ m/pixel. Both catalogs are then global and obtained from images with sufficient resolution to detect the lower limits of crater size mentioned before.  

The runs we selected went through two constraints: the mean fit with the observed MB populations and the formation of the two large basins in Vesta. However, due to small number statistics and high stochasticity of the model, we have stated earlier that the results in the individual runs for impactors larger than 10 km and, in consequence, craters larger than $\sim 100$ km may differ from run to run. Here we selected the run that we consider to be the best fit by comparing the simulated and cataloged distribution of craters larger than 100 km in Vesta, and we show the corresponding result of the craters on Ceres.

We plotted in Fig. \ref{fig:CraterTotalCV} the best fit for the total crater SFD for Ceres and Vesta, constructed by summing all the craters made by impactors from the different regions. We consider three different impact angles, $45\degr$, $90\degr$,  and $15\degr$, which are represented by a solid line and the top and bottom part of the shaded area, respectively. 

The craters discussed here were formed during the full 4 Gyr integration time of our runs. We do not consider cratering erasure processes such as cookie cutting, sandblasting, ejecta burial, or the effects of geometric saturation, nor any other geological processes that obliterate craters. Thus, the overprediction of craters is expected.

In the case of Ceres, we see that our model produces more large craters than what is currently observed. This is not surprising, as \cite{Hiesinger2016} and \citet{Marchi2016} show that Ceres is highly depleted in craters larger than $100$ km and lacks craters larger than $\sim 280$ km, and they attribute it to geological processes that obliterate large craters, in particular, as a result of Ceres’ internal processes. In fact, aqueous alteration processes and interior evolution processes have been suggested to explain the presence of water ice and organic material on the surface \citep{Prettyman2017, Desanctis2017} and those are important erosion mechanisms. Also, \cite{Ruesch2016} and \cite{Buczkowski2016} propose that Ceres has cryovolcanic activity that enabled resurfacing in geologically recent times. This overprediction is also consistent with low crater retention ages, which are estimated to be way shorter than 1 Gyr for the Kerwan region \citep{Hiesinger2016,Bottke2020}. We find that $\sim 13$  craters larger than $200$ km in Ceres over 4 Gyr should
have been formed. We also find the largest crater of $\sim 900 $ km, and the second largest one of $\sim 570$ km. This is remarkable considering that  \cite{Marchi2016} provide topographic evidence of a $\sim 800$ km diameter depression associated with an impact basin called Vendimia Planitia, and another one of $\sim 500$ km. If we add these two suggested basins to the SFD derived from \cite{gou2018}, we would find a good match with the current crater catalog in the large end.

Similarly, Fig. \ref{fig:CraterTotalCV} shows the total crater SFD for Vesta, along with the crater count derived from the catalog of \cite{Liu2018}. We see that our model is indeed able to reproduce the craters larger than 100 km, in agreement with the recent results of \cite{Roig2020}. In particular, the cratering retention age in Vesta has been estimated as smaller than 1.3 Gyr \citep{Bottke2020} for the Rheasilvia region. We find a clear overprediction in smaller sizes with respect to the cataloged craters. However, significant crater erasure took place in Vesta. In fact, the highest crater density is located in the northern hemisphere of Vesta, while the surface in the southern hemisphere was reset due to the impacts that formed the Rheasilvia and Veneneia basins, thus erasing the craters that were located there beforehand \citep{marchi2012,Vincent2014,Liu2018}. 

\subsection{Fragmentation}
Here we focus on the fragmentation of Ceres and Vesta. We have shown in the previous section that both bodies have been hit by a diverse collection of asteroids of many sizes coming from different source regions (Fig. \ref{fig:Impactors}). These impacts ejected fragments into the asteroid belt, according to Eqs. \ref{eq:MLF} and \ref{eq:pendiente}, and these fragments became new asteroids that further continued their collisional and dynamical evolution. The \texttt{ACDC} does not enable us to determine the final fate of the fragments that Ceres and Vesta ejected during their lives. Indeed, after the impact event, the fragments are part of the Inner or Middle belts, which are the regions where Vesta and Ceres are located, respectively, where they continue their collisional evolution according to their collisional and dynamical lifetimes. What this work can do is give us a hint as to all the material that was ejected from Ceres and Vesta during their history. In order to determine the final fate of these fragments, a full dynamical study must be performed. 

There are more small than big asteroids in all regions of the MB, and as we have shown, the same applies for the impactors. Therefore, smaller fragments are ejected in a more continuous way than bigger ones. Thus, we wish to estimate the production of small fragments today. To do so, using the second selection of runs, we summed the number of fragments ejected from Ceres and Vesta per million years during the last 10 Myr, and we derived median fragmentation rates per run. For Ceres, we obtain $\sim 126$ and $\sim 68\,700$ fragments per million years larger than $100$ m and $10$ m, respectively, and $\sim 31$ fragments larger than $1$ m per year. Similarly, for Vesta we obtain $\sim 30$ and $\sim 24\,400$ fragments per million years larger than $100$ m and $10$ m, respectively, and $\sim 14$ fragments larger than $1$ m per year. 

It is not possible to determine a fragmentation rate of bodies larger than 1 km due to the stochastic nature of our code. In fact, according to the scaling laws, these are produced when the impactors are larger than $\sim10$ km, which can hit Ceres and Vesta at any time during our simulations. This is more so the case for fragments larger than $10$ km, as they are created in isolated events that occur only once or twice per run when they are hit by the largest impactors. So, we are not able to provide an accurate estimation of the actual number of big fragments created in times comparable to their collisional lifetimes, as it greatly depends on the time of impact occurrence. Instead, for bodies larger than 1 km, we calculated the total number of fragments ejected during the whole 4 Gyr of evolution.

\begin{figure}
    \centering
    \includegraphics[width=8cm]{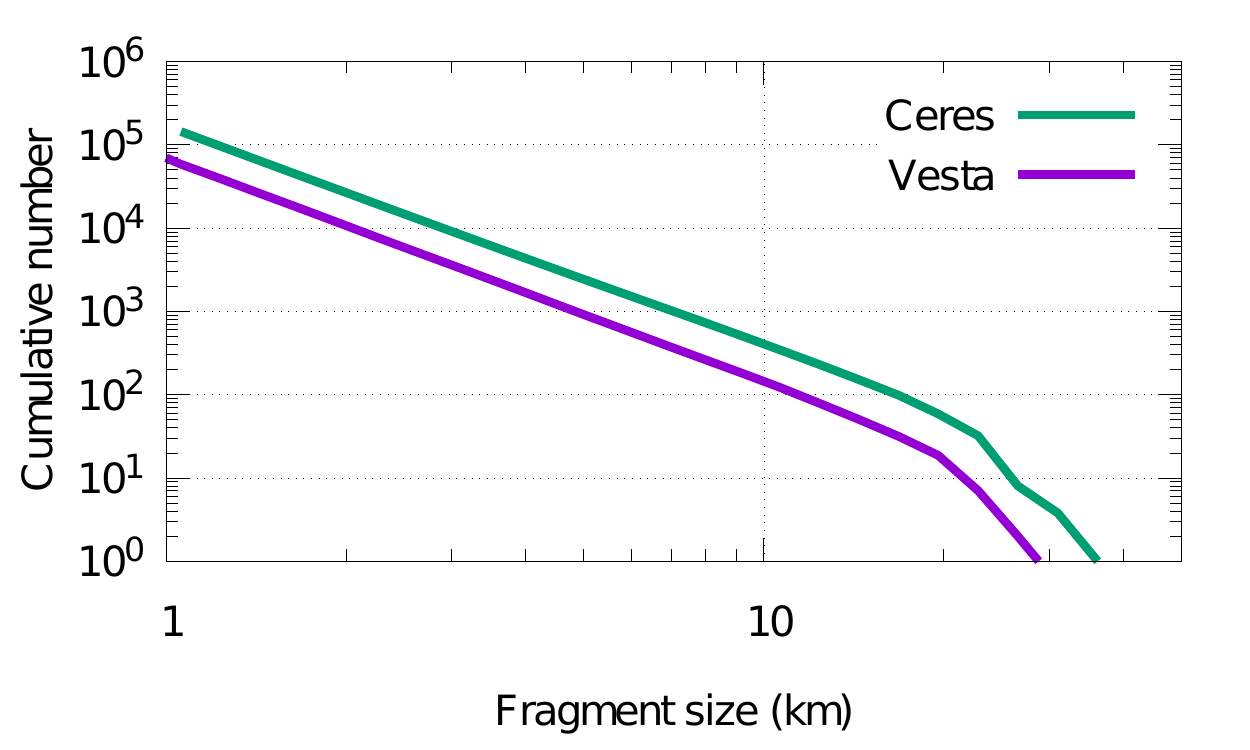}
    \caption{Median SFD of fragments larger than 1 km ejected from Ceres and Vesta during 4 Gyr of collisional evolution.}
    \label{fig:CV}
\end{figure}

The median SFD of fragments larger than $1$ km is shown in Figure \ref{fig:CV}.  We obtain $\sim 348$ and $\sim 123$ fragments larger than $10$ km, and the largest fragments of $\sim 36$ km and $\sim 26$ km for Ceres and Vesta, respectively.  However, since these fragments were ejected from both bodies over the age of the Solar System, it is very important to point out that this is not intended to be a SFD of a Ceres and Vesta family that we would expect to find today. In the case of Ceres, we see that a significant number of large asteroids should have been ejected from the body. However, an asteroid family for Ceres has not been identified \citep{Milani2014}, and the reason behind it is still unknown \citep{Rivkin2014,Carruba2016}. In order to determine the final fate of these fragments and thus explain the formation of the Vesta family and the absence of the Ceres family, and assuming no sublimation processes, a full dynamical study must be performed in the future performing $N$-body simulations of these bodies in times comparable to their collisional lifetimes. It is also very important to remark here that these fragments were obtained according to the scaling law we used in our collisional evolution model. In particular, as it was stated earlier, we used the scaling law derived by \cite{BenzAsphaug} for monolithic bodies made of basalt at $5$ km/s impact speeds for the entire MB. However, this implies the assumption that all bodies have the same composition and collide with the same velocities. It is very likely that there should be different scaling laws for different parts in the MB due to the diverse compositions of asteroids \citep{DeMeo2014}. However, as \cite{Cibulkova2014} show, scaling laws much different from those of \cite{BenzAsphaug} fail to reproduce the observed asteroid families. This is a very important matter to be addressed in future research. So, the results regarding the fragments of Ceres may change using a more appropriate scaling law, considering the amount of ice and volatiles in its surface, and the diversity of compositions in the MB.

\section{Conclusions}

In this work, we have studied the impacts on Ceres and Vesta using the \texttt{ACDC}, a six-part collisional evolution model of MB \citep{Zain2020}. The six regions of the MB we consider are the Inner, Middle, Pristine, Outer, Cybele, and High-Inclination belts. The \texttt{ACDC} is a statistical code that calculates the evolution in time of the number of objects in each part of the MB due to collisions between asteroids of the different regions, and it treats big impacts as random events. We used \texttt{ACDC} to determine what asteroids hit Ceres and Vesta and to ascertain where they came from, the craters they formed, and how many fragments were ejected into the MB. We summarize the results as follows:

\begin{itemize}
\item The six regions of the MB provide, to a greater or lesser extent, the impactors on Ceres and Vesta. In the case of Vesta, the relative contribution of the Inner, Middle, and Outer belt to the impactors is almost even. In the case of Ceres, the Outer belt is the main source of the impactors, which provides approximately half of the impactors smaller than $10$ km, followed by the Middle belt. The dominant taxonomic class in the Outer belt is the C-Type \citep{Demeo2013}, which is associated with carbonaceous objects and volatile-rich material. Therefore, a certain proportion of the water ice and organics present in Ceres could come from collisions received from the Outer belt. 
\item We were able to reproduce the craters larger than $100$ km in Vesta. In the case of Ceres, our runs represent the formation of two large depressions present in Ceres associated with possible impact basins \citep{Marchi2016}. These constitute the largest crater of $\sim 900$ km and the second largest one of $\sim 570$ km. 
\item Throughout their collisional history, Ceres and Vesta ejected fragments into the asteroid belt. We obtain fragmentation rates of tens of fragments larger than $1$ m per year to tens of fragments larger than $100$ m per million years for Vesta and a factor of $\sim 4$ greater for Ceres. In larger sizes, due to the high stochasticity of impact events, we obtain the total number of fragments ejected through $4$ Gyr of collisional evolution. We find that hundreds of bodies larger than $10$ km should have been ejected from Ceres and Vesta during their history. However, this work did not enable us to determine the final fate of these fragments, which is relevant to explain the absence of a Ceres asteroid family or the formation of the current Vesta family. In order to do so, a full dynamical study must be performed, using $N$-body simulations to determine the evolution of these fragments. The use of more appropriate scaling laws in collisional evolution models, accounting for the diversity of compositions in the MB, may also have an impact on these results regarding the fragmentation of asteroids in the MB. These are very relevant topics to be reviewed in future work.
\end{itemize}

The use of an improved collisional evolution model that accounts for different dynamical features in the MB, such as the one performed in this work, is undoubtedly a constructive contribution to the understanding of the collisional history of Ceres and Vesta. We consider the results and controversies discussed in this paper as interesting starting points for further research involving Ceres, Vesta, and the collisional and dynamical evolution of other minor bodies in the MB and the Solar System.

\begin{acknowledgements}
The present investigation work was partially financed by Agencia Nacional de Promoci\'on Cient\'ifica y Tecnol\'ogica (ANPCyT) through PICT 201-0505, and by Universidad Nacional de La Plata (UNLP) through PID G144. We acknowledge the financial support by Facultad de Ciencias Astron\'omicas y Geof\'isicas de La Plata (FCAGLP) and Instituto de Astrof\'isica de La Plata (IALP) for extensive use of their computing resources. We wish to thank Sheng Gou and Zongyu Yue for kindly sharing cratering data of Ceres and Vesta with us.
\end{acknowledgements}

\bibliographystyle{aa} % style aa.bst
\bibliography{Zain2021} % your references Yourfile.bib

\end{document}